\documentclass[12pt]{extarticle}
\usepackage{tikz}
\usepackage{fleqn}
\usepackage{graphicx}
\usepackage{amsfonts,amsthm}
\usepackage{amsthm} 
\makeatletter
\newcommand*{\encircled}[1]{\relax\ifmmode\mathpalette\@encircled@math{#1}\else\@encircled{#1}\fi}
\newcommand*{\@encircled@math}[2]{\@encircled{$\m@th#1#2$}}
\newcommand*{\@encircled}[1]{%
  \tikz[baseline,anchor=base]{\node[draw,circle,outer sep=0pt,inner sep=.2ex] {#1};}}
\makeatother
\usepackage{amsmath}
\usepackage{amssymb}
\begin{document}
\Large
\begin{center}
\bf{Taxonomy of Polar Subspaces of Multi-Qubit Symplectic Polar Spaces of Small Rank}
\end{center}
\vspace*{-.3cm}

\large
\begin{center}
 Metod Saniga$^{1}$, Henri de Boutray$^{2}$, Fr\'ed\'eric Holweck$^{3}$ and\\ Alain Giorgetti$^{2}$
\end{center}
\vspace*{-.6cm} 

\normalsize
\begin{center}

$^{1}$Astronomical Institute of the Slovak Academy of Sciences,\\
SK-05960 Tatransk\' a Lomnica, Slovak Republic\\
(msaniga@astro.sk)
\vspace*{0.1cm}

$^{2}$Institut FEMTO-ST, DISC $-$ UFR-ST, Universit\'e Bourgogne Franche-Comt\'e, F-25030 Besan\c con, France\\ 
(henri.deboutray@femto-st.fr, alain.giorgetti@femto-st.fr)
\vspace*{0.0cm}

and
\vspace*{0.0cm}

$^{3}$Laboratoire Interdisciplinaire Carnot de Bourgogne, ICB/UTBM, UMR 6303 CNRS,
Universit\'e Bourgogne Franche-Comt\'e, F-90010 Belfort, France\\ (frederic.holweck@utbm.fr)

\end{center}

\vspace*{-.4cm} \noindent \hrulefill

\vspace*{-.1cm} \noindent {\bf Abstract}

\noindent
We study certain physically-relevant subgeometries of binary symplectic polar spaces $W(2N-1,2)$ of small rank $N$, when the points of these spaces canonically encode $N$-qubit observables. Key characteristics of a subspace of such a space $W(2N-1,2)$ are: the number of its negative lines, the distribution of types of observables, the character of the geometric hyperplane the subspace shares with the distinguished (non-singular) quadric of $W(2N-1,2)$ and the structure of its Veldkamp space. In particular, we classify and count polar subspaces of $W(2N-1,2)$ whose rank is $N-1$.  $W(3,2)$ features three negative lines of the same type and its $W(1,2)$'s are of five different types. $W(5,2)$ is endowed with 90 negative lines of two types and its $W(3,2)$'s split into 13 types. 279 out of 480 $W(3,2)$'s with three negative lines are composite, i.\,e. they all originate from the two-qubit $W(3,2)$. Given a three-qubit $W(3,2)$ and any of its geometric hyperplanes, there are three other $W(3,2)$'s possessing the same hyperplane. The same holds if a geometric hyperplane is replaced by a `planar' tricentric triad. A hyperbolic quadric of $W(5,2)$ is found to host particular sets of seven $W(3,2)$'s, each of them being uniquely tied to a Conwell heptad with respect to the quadric. 
There is also a particular type of $W(3,2)$'s, a representative of which features a point each line through which is negative. Finally, $W(7,2)$ is found to possess 1908 negative lines of five types and its $W(5,2)$'s fall into as many as 29
types. 1524 out of 1560 $W(5,2)$'s with 90 negative lines originate from the three-qubit $W(5,2)$. 
Remarkably, the difference in the number of negative lines for any two distinct types of four-qubit $W(5,2)$'s is a multiple of four.     
\\

\vspace*{-.2cm}
Distinguished Sets of Doilies -- Geometric Hyperplanes -- Veldkamp lines 


\vspace*{-.2cm} \noindent \hrulefill

\noindent
\section{Introduction}
Some fifteen years ago it was discovered (see, e.g., \cite{sapla,plasa,hos,kthas}) that there exists a deep connection between the structure of the $N$-qubit Pauli group and that of the binary symplectic polar space of rank $N$, $W(2N-1,2)$, where commutation relations between elements of the group are encoded in collinearity relations between points of $W(2N-1,2)$. This connection has subsequently been used 
to get a deeper insight into, for example, finite geometric nature of observable-based proofs of quantum contextuality (for a recent review, see \cite{holw}), properties of certain black-hole entropy formulas \cite{lsvp} and the so-called black-hole/qubit correspondence \cite{bdl}, leading to finite-geometric underpinning four distinct Hitchin's invariants and the Cartan invariant of form theories of gravity \cite{lhs} and even to an intriguing finite-geometric toy model of space-time~\cite{lehol}. This group-geometric connection was further strengthened by making use of the concept of geometric hyperplane and that of the Veldkamp space of $W(2N-1,2)$ \cite{vl}. As per quantum contextuality, famous two-qubit Mermin-Peres magic squares were found to be isomorphic to a special class of geometric hyperplanes of $W(3,2)$ called grids \cite{spph}, whereas three-qubit Mermin pentagrams were found to have their natural settings in the magic Veldkamp line of $W(5,2)$ \cite{levsza}, being also isomorphic -- under the grassmannian correspondence of type Gr$(2,4)$ -- to ovoids of $W(3,2)$ \cite{salev}. Concerning the black-hole/qubit correspondence, here a key role is played by the geometric hyperplane isomorphic to an elliptic quadric of $W(5,2)$. Interestingly, form theories of gravity seem to indicate that a certain part of the magic Veldkamp line in the four-qubit symplectic polar space, $W(7,2)$, and the associated extended geometric hyperplanes are of physical relevance as well.

From the preceding paragraph it is obvious that revealing finer traits of the structure of binary symplectic polar spaces of small rank can be vital for further physical applications  of these spaces. Having this in view, we  will focus on sets of $W(2N-3,2)$'s located in $W(2N-1,2)$, for $N= 2, 3, 4$, providing their comprehensive observable-based taxonomy. 
Key parameters of our classification of such subspaces of $W(2N-1,2)$ will be: the number of negative lines they contain (which is also an important parameter when it comes to quantum contextuality), the distribution of different types of observables they feature, the character of the geometric hyperplane a subspace of a given type shares with the distinguished (non-singular) quadric of $W(2N-1,2)$ and, in the case of refined `decomposition' of three-qubit $W(3,2)$'s, also the very structure of their Veldkamp lines.

The paper is organized as follows. Sec.\,2 provides the reader with the necessary finite-geometric background and notation. Sec.\,3 deals with $W(3,2)$ and the hierarchy of its triads. Sec.\,4 addresses the three-qubit $W(5,2)$ and its $W(3,2)$'s; here we classify $W(3,2)$'s in two distinct ways and illustrate the fact that there are four $W(3,2)$'s sharing a geometric hyperplane, or a specific tricentric triad. Sec.\,5 focuses on prominent septuplets of $W(3,2)$'s that are closely related to Conwell heptads with respect to a hyperbolic quadric of $W(5,2)$. Sec.\,6 classifies $W(5,2)$'s living in the four-qubit $W(7,2)$ and furnishes a couple of examples of their composite types. Finally, Sec.\,7 is devoted to concluding remarks.

\section{Finite Geometry Background}
Given a $d$-dimensional projective space PG$(d,2)$ over GF$(2)$,  
a {\it polar space} ${\cal P}$ in this projective space consists of 
the projective subspaces that are {\it totally isotropic/singular} with respect to 
a given non-singular bilinear form; PG$(d,2)$ is called the {\it ambient projective space} of ${\cal P}$. 
A projective subspace of maximal dimension in ${\cal P}$ is called a {\it generator}; 
all generators have the same (projective) dimension $r - 1$. 
One calls $r$ the {\it rank} of the polar space.

Polar spaces of relevance for us are of three types (see, for example, \cite{hita,cam}): symplectic, hyperbolic and elliptic.
The {\it symplectic} polar space $W(2N - 1,2)$, $N \geq 1$, consists of all the points of PG$(2N - 1, 2)$,
$\{(x_1, x_2, \ldots, x_{2N}): x_j \in \{0,1\}, j \in \{1,2, \ldots, 2N \}\}\backslash \{(0, 0, \ldots, 0)\}$,
 together with the totally isotropic subspaces with respect to the standard symplectic form
\begin{equation} 
\sigma(x,y) = x_1 y_{N+1} - x_{N+1} y_1 +  x_2 y_{N+2} - x_{N+2} y_2 + \dots + x_{N} y_{2N} - x_{2N} y_{N}.
\label{symplf}
\end{equation}
This space features
\begin{equation}
|W|_p = 4^N - 1
\label{ptsinwn}
\end{equation}
points and
\begin{equation}
|W|_g = (2+1)(2^2+1)\cdots(2^N+1)
\label{geninwn}
\end{equation}
generators.
The {\it hyperbolic} orthogonal polar space $\mathcal{Q}^{+}(2N - 1, 2)$, $N \geq 1$, 
is formed by all the subspaces of  PG$(2N - 1, 2)$ that lie on a given non-singular hyperbolic quadric, with the standard equation 
\begin{equation}
x_1 x_{N+1} + x_2 x_{N+2} \ldots + x_{N} x_{2N} = 0.
\label{hqsteqn}
\end{equation}
Each $\mathcal{Q}^{+}(2N - 1, 2)$ contains 
\begin{equation}
|Q^{+}|_p = (2^{N-1} + 1)(2^{N} -1) 
\label{ptsonhq}
\end{equation}
points and there are 
\begin{equation}
|W|_{Q^{+}} = |Q^{+}|_p + 1 = (2^{N-1} + 1)(2^{N} -1) + 1
\label{hqinwn}
\end{equation}
copies of them in $W(2N - 1,2)$.
Finally, the {\it elliptic} orthogonal polar space $\mathcal{Q}^{-}(2N - 1,2)$, $N \geq 2$,
comprises all points and subspaces of PG$(2N - 1, 2)$ satisfying the standard equation 
\begin{equation}
f(x_1,x_{N+1})+x_2 x_{N+2}+\cdots+x_{N}x_{2N} = 0, 
\label{eqsteqn}
\end{equation}
where $f$ is an irreducible polynomial over GF$(2)$.
Each $\mathcal{Q}^{-}(2N - 1, 2)$ contains 
\begin{equation}
|Q^{-}|_p = (2^{N-1} - 1)(2^{N} + 1) 
\label{ptsoneq}
\end{equation}
points and there are 
\begin{equation}
|W|_{Q^{-}} = |Q^{-}|_p + 1 = (2^{N-1} - 1)(2^{N} + 1) + 1
\label{eqinwn}
\end{equation}
copies of them in $W(2N - 1,2)$.
For both symplectic and hyperbolic polar spaces $r = N$, whereas for the elliptic one $r = N - 1$.
The smallest non-trivial symplectic polar space is the $N=2$ one, $W(3,2)$, often referred to as the {\it doily}.
It features 15 points (see eq.\,(2)) and the same number of lines (that are also its generators, see eq.\,(\ref{geninwn})), with three points
per line and three lines through a point; it is a self-dual $15_3$-configuration and the only 
one out of 245\,342 such configurations that is triangle-free, being, in fact, isomorphic to the generalized quadrangle of order two (GQ$(2,2)$). This symplectic polar space features ten $\mathcal{Q}^{+}(3, 2)$'s
(by eq.\,(\ref{hqinwn})) and six $\mathcal{Q}^{-}(3, 2)$'s (by eq.\,(\ref{eqinwn})). A $\mathcal{Q}^{+}(3, 2)$ contains nine points and six lines forming a $3 \times 3$ grid, so it is also called a grid. A $\mathcal{Q}^{-}(3, 2)$ features five pairwise non-collinear points, hence it is an ovoid. A triple of mutually non-collinear points of $W(3,2)$ is called a {\it triad} and a point collinear with all the three points of a triad is called a {\it center} of the triad; $W(3,2)$ contains 60 unicentric and 20 tricentric triads.   

The $N$-qubit observables we will be dealing with belong to the set
\begin{equation}
{\cal S}_N = \{G_1 \otimes G_2 \otimes \cdots \otimes G_N:~ G_j \in \{I, X, Y, Z \},~ j \in \{1, 2, \ldots, N \}\} \backslash 
\{{\cal I}_N \}
\label{nqobs}
\end{equation}
where ${\cal I}_N \equiv I_{(1)} \otimes I_{(2)} \otimes \ldots \otimes I_{(N)}$, 
\begin{equation}
X = \left(
\begin{array}{rr}
0 & 1 \\
1 & 0 \\
\end{array}
\right),~~
Y = \left(
\begin{array}{rr}
0 & -i \\
i & 0 \\
\end{array}
\right),~~{\rm and}~~
Z = \left(
\begin{array}{rr}
1 & 0 \\
0 & -1 \\
\end{array}
\right)
\label{paulis}
\end{equation}
are the Pauli matrices, $I$ is the identity matrix and `$\otimes$' stands for the tensor
product of matrices. ${\cal S}_N$, whose elements are simply those of the $N$-qubit Pauli group if the global phase is disregarded, features two kinds of observables, namely {\it symmetric} (i.e., observables featuring an even number of $Y$'s) and {\it skew-symmetric}; the number of symmetric observables is $(2^{N-1} + 1)(2^{N} -1)$. We shall further employ
a finer classification where an observable having $N-1$, $N-2$, $N-3$, \dots $I$'s will be, respectively, of type $A$, $B$, $C$,
\dots; also, whenever it is clear from the context, $G_1 \otimes G_2 \otimes \cdots \otimes G_N$ will be short-handed to
$G_1 G_2 \cdots G_N$.

For a particular value of $N$, 
the $4^N - 1$ elements of ${\cal S}_N$ can be bijectively identified with the same number of points of
$W(2N-1, 2)$ in such a way that the images of two commuting elements lie on \underline{the same} line of this polar space, and
{\it generators} of $W(2N - 1, 2)$ correspond to \underline{maximal} sets of mutually commuting elements.
If we take the symplectic form defined by eq. (1), then this bijection acquires the form
\begin{equation}
G_j \leftrightarrow (x_j, x_{j+N}),~j \in \{1, 2,\dots,N\},
\label{nobspts}
\end{equation}
assuming that
\begin{equation}
I \leftrightarrow (0,0),~X \leftrightarrow (0,1),~Y \leftrightarrow (1,1),~ {\rm and}~Z \leftrightarrow (1,0).
\label{paulipts}
\end{equation}
Employing the above-introduced bijection (for more details see, e.\,g., \cite{levsza}), it can be shown that given an observable $O$,  the set of symmetric observables commuting with $O$ together with the set of skew-symmetric observables not
commuting with $O$ will lie on a certain non-degenerate quadric of  $W(2N - 1,2)$, this quadric being hyperbolic (resp. elliptic) if $O$ is symmetric (resp. skew-symmetric). We can express this important property by making, whenever appropriate, this associated
observable explicit in a subscript, $\mathcal{Q}^{\pm}_{(O)}(2N - 1, 2)$, noting that there exists a particular hyperbolic quadric associated with ${\cal I}$:

\begin{equation}
 \begin{split}
\mathcal{Q}^{+}_{({\cal I})}(2N - 1, 2) := \{(x_1, x_2, \dots, x_{2N}) \in W(2N-1,2)~|~ x_1 x_{N+1} + x_2 x_{N+2} + \\
\ldots + x_{N} x_{2N} = 0 \}.
 \end{split}
\label{ndisthq}
\end{equation}

 Given a point-line incidence geometry $\Gamma(P, L)$, a {\it geometric hyperplane} of $\Gamma(P, L)$ is a subset of its point set such that a line of the geometry is either {\it fully} contained in the subset or has with it just a {\it single} point in common. The {\it Veldkamp} space $\mathcal{V}(\Gamma)$ of $\Gamma(P, L)$  is the
space in  which \cite{buco}: (i) a point is a geometric hyperplane of  $\Gamma$ and (ii) a line is the collection, denoted $H'H''$, of all geometric
hyperplanes $H$ of $\Gamma$  such that $H' \cap H'' = H' \cap H = H'' \cap H$ or $H = H', H''$, where $H'$ and $H''$ are distinct points of $\mathcal{V}(\Gamma)$. For a $\Gamma(P, L)$ with \underline{three} points on a line, all Veldkamp lines are of the form
$\{H', H'', \overline{H' \Delta H''}\}$ where $\overline{H' \Delta H''}$
is the complement of symmetric difference of $H'$ and $H''$, i.\,e. they form a
vector space over GF$(2)$. As demonstrated in \cite{vl}, $\mathcal{V}(W(2N-1,2)) \cong $ PG$(2N, 2)$. Its points are both hyperbolic and elliptic quadrics
of $W(2N-1,2)$, as well as its perp-sets. Given a point $x$ of $W(2N-1,2)$, the {\it perp-set}   
$\widehat{\mathcal{Q}}_{(x)}(2N-1,2)$ of $x$ consists of all the points collinear with it, 
\begin{equation}
\widehat{\mathcal{Q}}_{(x)}(2N-1,2):= \{y \in W(2N-1,2)~|~ \sigma(x,y) = 0 \};
\label{nperp}
\end{equation} 
the point $x$ being referred to as the {\it nucleus} of $\widehat{\mathcal{Q}}_{(x)}(2N-1,2)$.

\begin{figure}[t]
\centerline{\includegraphics[width=9truecm,clip=]{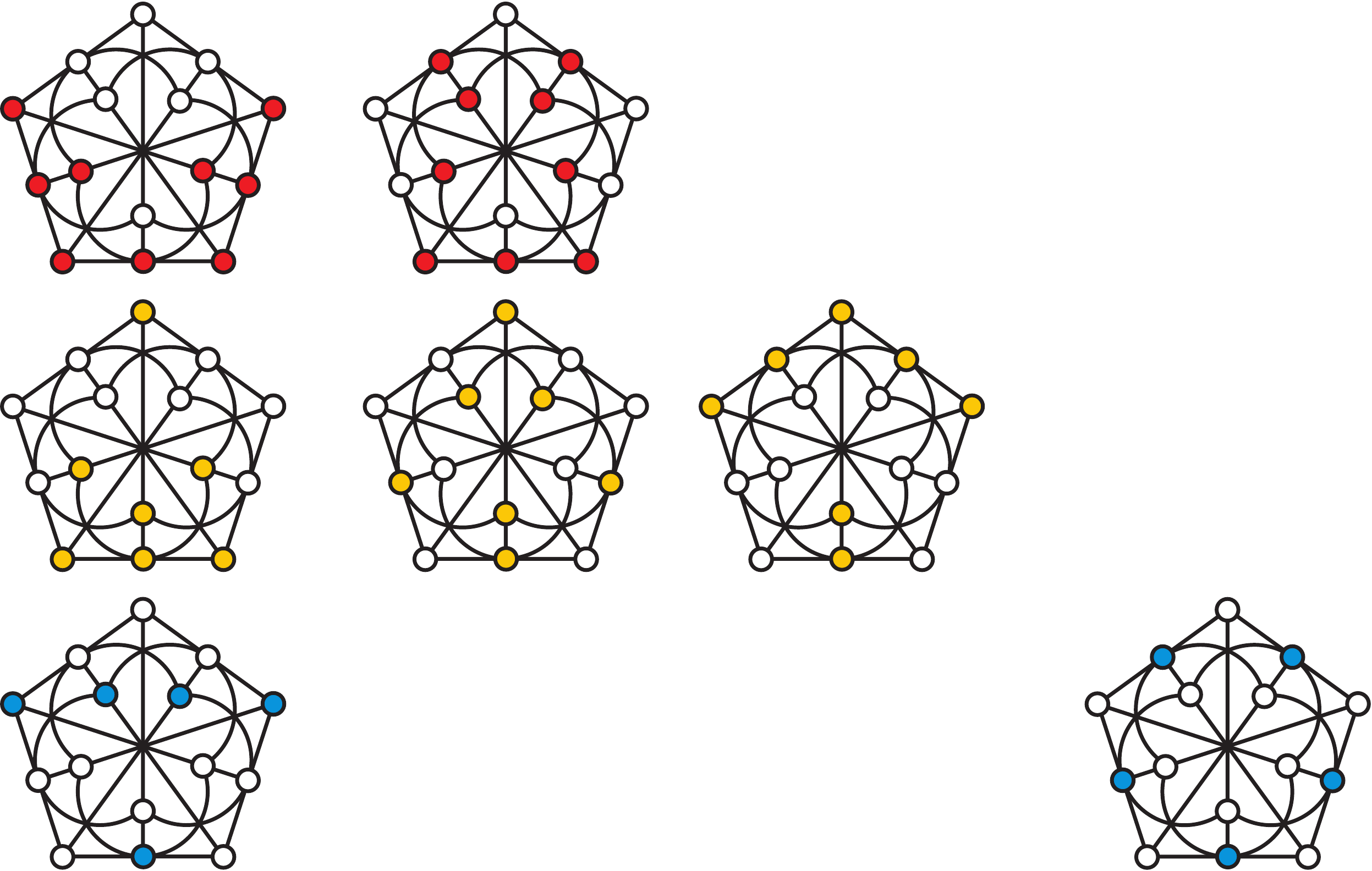}}
\vspace*{.2cm}
\caption{The three kinds of geometric hyperplanes of $W(3,2)$. The 15 points of the doily are represented by small circles and its 15 lines are illustrated by the straight segments as
well as by the segments of circles; note that not every intersection of two segments counts for a point of
the doily. The upper panel shows grids (red bullets), the middle panel perp-sets (yellow bullets) and the bottom panel ovoids (blue bullets). Each picture $-$ except that located in the bottom right-hand corner $-$
stands for five different hyperplanes, the four other being obtained from it by its successive rotations through 72 degrees around the center of the pentagon.}
\label{ill2qghs}
\end{figure}
\begin{table}[pth!]
\begin{center}
\caption{An overview of the properties of the five different types of lines of $\mathcal{V}(W(3,2))$
in terms of the {\it core} (i.\,e., the set of points common to all the three hyperplanes forming the line) and the types of geometric hyperplanes featured by a generic line of a given type.
The last column gives the total number of lines per each type.}
\label{t2qubvls}
 \vspace*{0.4cm}
\begin{tabular}{||c|c|ccc|c||}
\hline \hline
Type & Core & Perps & Ovoids & Grids & $\#$ \\
\hline
I & Two Secant Lines & 1 & 0 & 2 & 45\\
II & Single Line & 3 & 0 & 0 & 15 \\
III &Tricentric Triad & 3 & 0 &  0 & 20 \\
IV & Unicentric Triad & 1 & 1 &  1 & 60 \\
V & Single Point & 1 &  2 & 0 & 15 \\
\hline \hline
\end{tabular}
\end{center}
\end{table}

\begin{figure}[t]
\centerline{\includegraphics[width=6.0truecm,clip=]{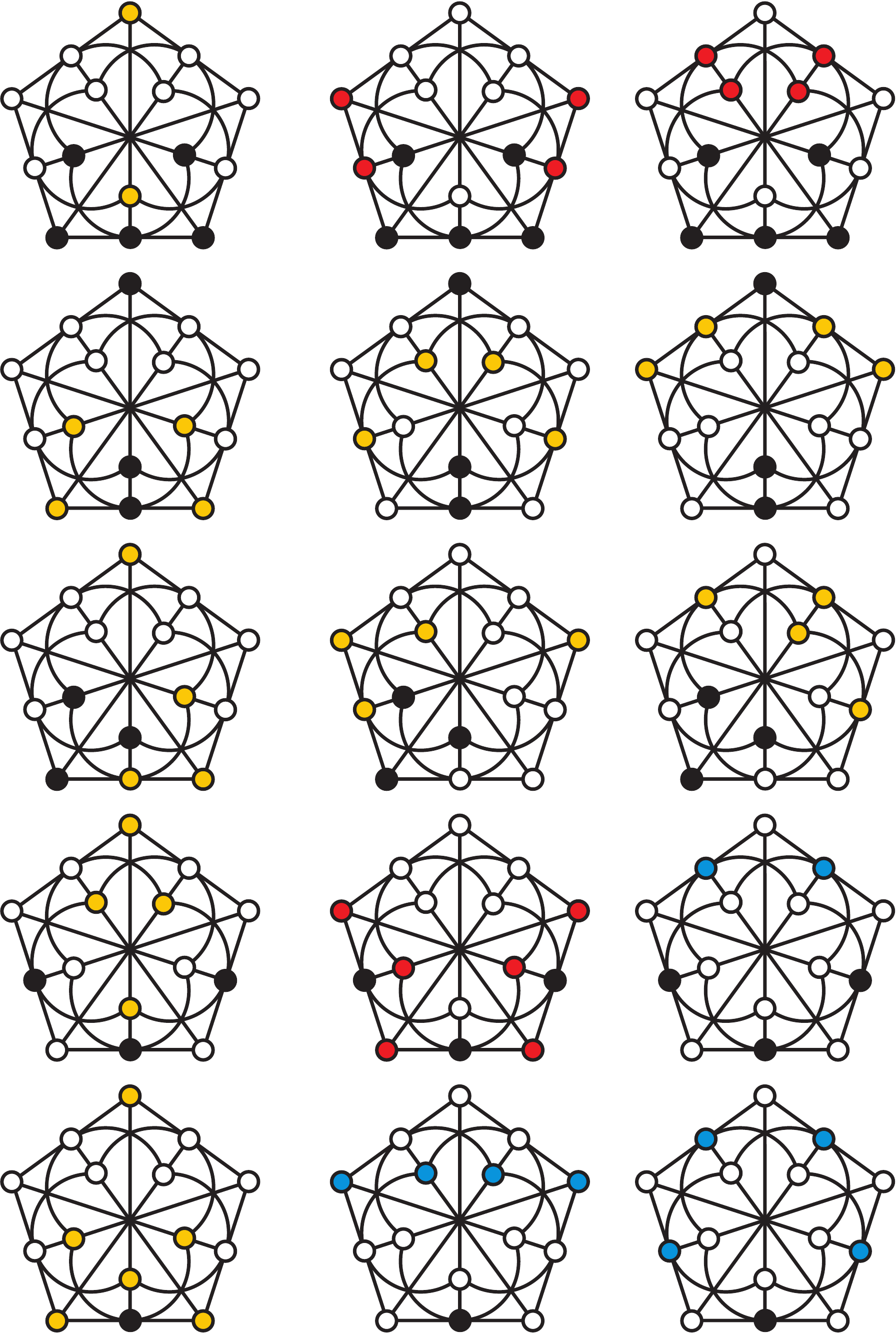}}
\vspace*{.2cm} \caption{An illustrative portrayal of representatives (rows) of the five (numbered consecutively from top to bottom)  different types of lines of
$\mathcal{V}(W(3,2))$, each being uniquely determined by the properties of its core (black bullets).}
\label{ill2qvls}
\end{figure}

We shall briefly recall basic properties of the Veldkamp space of the doily, $\mathcal{V}(W(3,2)) \simeq$ PG(4,\,2), whose in-depth description can be found in \cite{spph}. 
The 31 points of $\mathcal{V}(W(3,2))$ comprise 15 perp-sets, ten grids  and six ovoids $-$ as also illustrated in Figure \ref{ill2qghs}.
The 155 lines of $\mathcal{V}(W(3,2))$ split into five distinct types as summarized in Table \ref{t2qubvls} and depicted in Figure \ref{ill2qvls}.
(Table \ref{t2qubvls}, as well as Figure \ref{ill2qghs} and Figure \ref{ill2qvls}, were taken from \cite{slpp}.)

In what follows we will mainly be focused on $W(2N - 3, 2)$'s that are located in $W(2N - 1, 2)$.
These are, in general, of two different kinds \cite{vl}.  A $W(2N - 3, 2)$ of the first kind, to be called {\it linear},  is isomorphic to the intersection of two perp-sets with non-collinear nuclei and their number in $W(2N - 1, 2)$ is
\begin{equation}
|W|_{W_l} = \frac{1}{3} 4^{N-1}(4^N - 1).
\label{linwinwn}
\end{equation}
A $W(2N - 3, 2)$ of the second kind, to be called {\it quadratic}, is isomorphic to the intersection of a hyperbolic quadric and an elliptic quadric and $W(2N - 1, 2)$ features 
\begin{equation}
|W|_{W_q} = 4^{N-1}(4^N - 1)
\label{quadwinwn}
\end{equation}
of them. By way of example, in $W(3,2)$ a linear (resp. quadratic) $W(1,2)$ corresponds to a tricentric (resp. unicentric) triad. 

In the sequel, when referring to $W(2N-1,2)$ and its subspaces we will always have in mind the $W(2N-1,2)$ and its subspaces whose
points are labelled by $N$-qubit observables from the set ${\cal S}_N$ as expressed by eqs. (\ref{nobspts}) and 
(\ref{paulipts}). 
Moreover, a linear subspace of such $W(2N-1,2)$ will be called {\it positive} or {\it negative} according as the (ordinary) product of
the observables located in it is $+ {\cal I}_N$ or $- {\cal I}_N$, respectively. 
Let us illustrate this point taking again the $N=2$ case. Up to isomorphism, there is just one type of the two-qubit
doily. Its six observables of type $A$ are $IX$, $XI$, $IY$, $YI$, $IZ$, and $ZI$ and its nine ones of type $B$ are
$XX$, $XY$, $XZ$, $YX$, $YY$, $YZ$, $ZX$, $ZY$ and $ZZ$, the latter lying on a particular hyperbolic quadric, 
$\mathcal{Q}_{(YY)}^{+}(3, 2)$. Among the 15 lines only the three lines $\{XX, YY, ZZ\}$, $\{XY, YZ, ZX\}$ and
$\{XZ, YX, ZY\}$ are negative, forming also one system of generators of $\mathcal{Q}_{(YY)}^{+}(3, 2)$.

\section{$\boldsymbol{W(3,2)}$ and its Two-Qubit $\boldsymbol{W(1,2)}$'s}
This is a rather trivial case. As already mentioned in Sec. 2, the doily contains three negative lines, of the same ($B-B-B$) type. Among its $W(1,2)$'s, we find two types of linear ones and three types of quadratic ones, whose properties are summarized in Table \ref{tw1inw3}.

\begin{table}[h]
\begin{center}
\caption{Classification of $W(1,2)$'s living in $W(3,2)$. Column one ($T$) shows the type, columns two and three ($O_{A}$ and $O_{B}$) indicate the number of observables of corresponding types, and columns four ($W_{l}$) and five ($W_{q}$) yield, respectively, the number of `linear' and `quadratic' $W(1,2)$'s  of a given type. } 
\label{tw1inw3}
\vspace*{0.4cm}
\scalebox{0.8}{
\begin{tabular}{|r|cc|rr|}
\hline \hline
$T$        & $O_{A}$ & $O_{B}$ & $W_{l}$  & $W_{q}$    \\
\hline 
  1        & 0       & 3       &  $-$     & 6            \\
  2        & 1       & 2       &  $-$     & 36            \\	
  3        & 1       & 2       &  18      & $-$            \\	
	4        & 2       & 1       &  $-$     & 18            \\	
	5        & 3       & 0       &  2       & $-$            \\	
\hline \hline
\end{tabular}
}
\end{center}
\end{table}
It is worth noticing that the six quadratic $W(1,2)$'s (i.e., unicentric triads) of Type 1 lie on the distinguished quadric
$\mathcal{Q}_{(YY)}^{+}(3, 2)$, being in fact its six ovoids.

\section{$\boldsymbol{W(5,2)}$ and its Three-Qubit Doilies}

The space $W(5,2)$ contains 63 points, 315 lines and 135 generators, the latter being all Fano planes. Among the 63 canonical three-qubit
observables associated to the points, nine are of type $A$, 27 are type $B$ and 27 are of type $C$. Through an observable of type $C$ there pass six
negative lines, all being of type $C-C-B$; the total number of negative lines of this type thus is  
$ \frac{27 \times 6}{2} = 81.$
Through an observable of type $B$ there pass four negative lines. Of them, three are of the above-mentioned type and the fourth
one is of type $B-B-B$; the total number of negative lines of the latter type is
$\frac{27 \times 1}{3} = 9.$
As no negative line features an observable of type $A$, one finds that the $W(5,2)$ accommodates as many as $(81+9=)~90$ negative lines. 

\begin{figure}[pth!]
\centerline{\includegraphics[width=3.5cm,clip=]{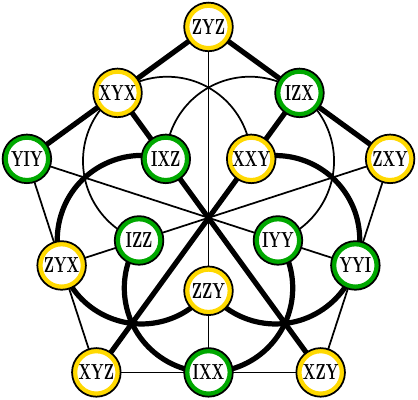}\hspace*{1.9cm}\includegraphics[width=3.5cm,clip=]{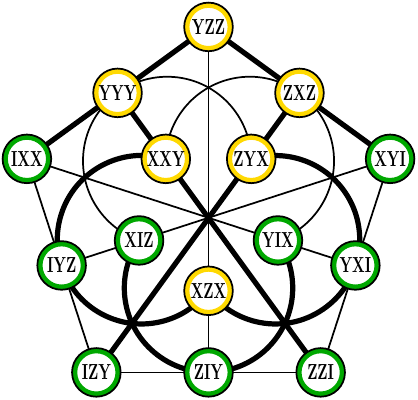}
\hspace*{1.7cm}\includegraphics[width=3.5cm,clip=]{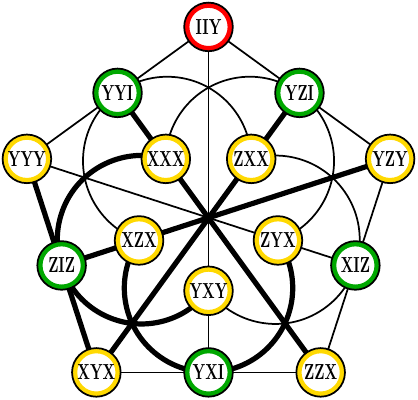}}

\vspace*{.6cm}

\centerline{\includegraphics[width=3.5cm,clip=]{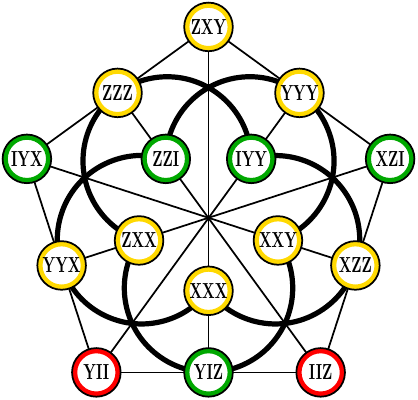}\hspace*{1.9cm}\includegraphics[width=3.5cm,clip=]{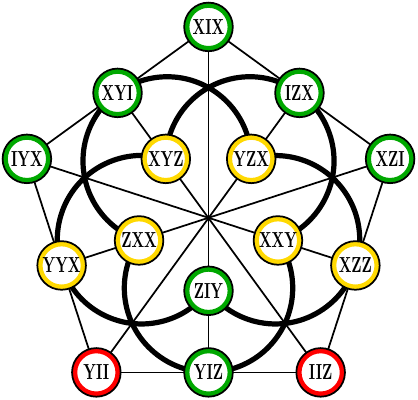}
\hspace*{1.7cm}\includegraphics[width=3.5cm,clip=]{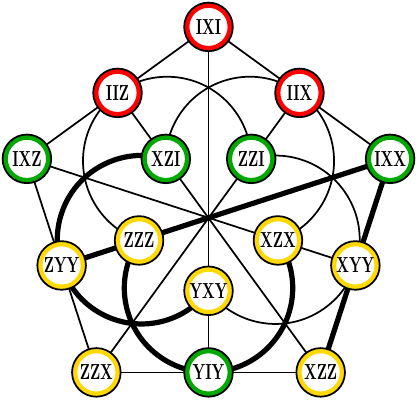}}

\vspace*{.6cm}

\centerline{\includegraphics[width=3.5cm,clip=]{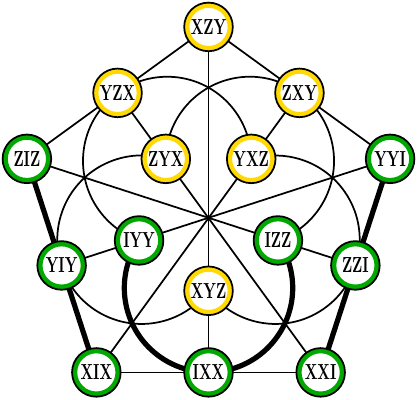}\hspace*{1.9cm}\includegraphics[width=3.5cm,clip=]{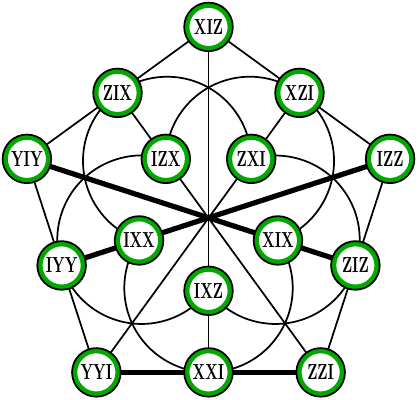}
\hspace*{1.7cm}\includegraphics[width=3.5cm,clip=]{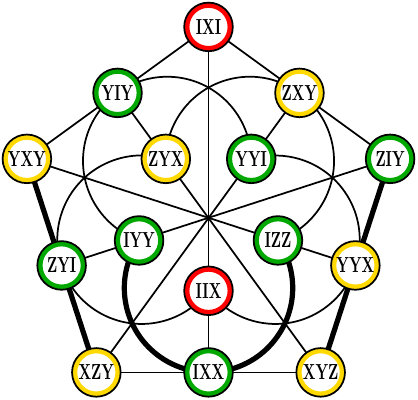}}

\vspace*{.6cm}

\centerline{\includegraphics[width=3.5cm,clip=]{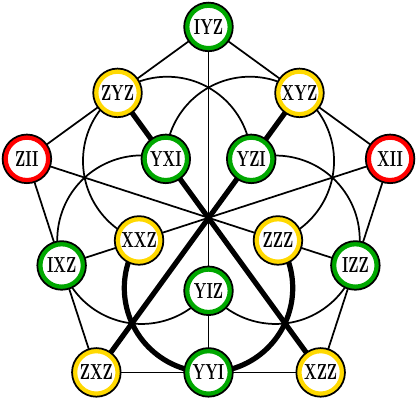}\hspace*{1.9cm}\includegraphics[width=3.5cm,clip=]{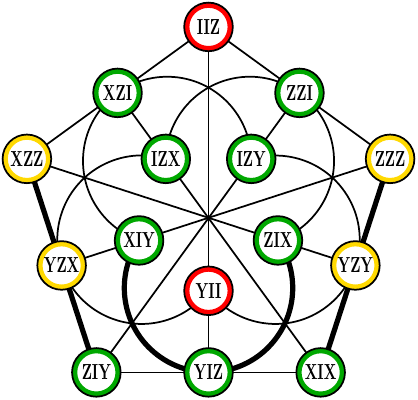}
\hspace*{1.7cm}\includegraphics[width=3.5cm,clip=]{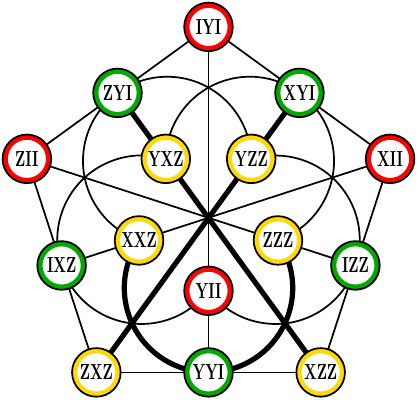}}

\vspace*{.6cm}

\centerline{\includegraphics[width=3.5cm,clip=]{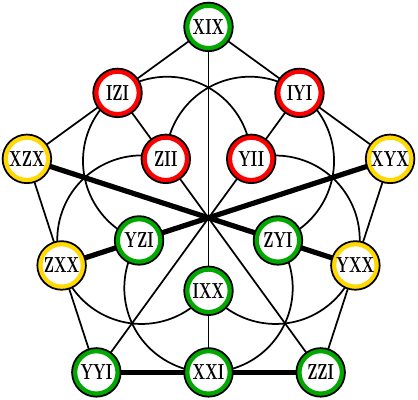}\hspace*{1.9cm}\includegraphics[width=3.5cm,clip=]{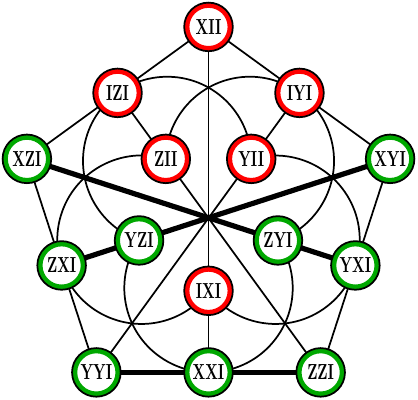}
\hspace*{5.2cm}}
\caption{Representatives -- numbered consecutively from left to right, top to bottom -- of the 13 different types of three-qubit doilies; Type 1 is top left, Type 13 bottom middle; we also distinguish between Type $9A$ (3rd row, right) and Type $9B$ (4th row, left). The three different types of observables are distinguished by different colors and the negative lines are drawn heavy.}
\label{ill3qdoilies}
\end{figure}

When we pass to $W(3,2)$'s, we find a (much) richer structure, because alongside the types of observables we can employ
one more parameter, namely the number of negative lines a given $W(3,2)$ contains. In fact, we find that the 336 linear doilies
(see eq.\,\ref{linwinwn}) fall into six different types and the 1008 quadratic ones (see eq.\,\ref{quadwinwn}) into seven types; we note in passing that  Type 9 splits further into two subtypes depending on whether the two observables of type $A$ do  (Type $9A$, 162 members) or do not (Type $9B$, 54 members) commute. 
This classification is summarized in Table \ref{tw3inw5} and also pictorially illustrated in Figure \ref{ill3qdoilies}.
It is worth noticing here that there are two different types of doilies (Type 3 and Type 6) exhibiting an even number of negative lines.

\begin{table}[t]
\begin{center}
\caption{Classification of doilies living in $W(5,2)$. Column one ($T$) shows the type, column two ($C^{-}$) the number of negative lines in a doily of the given type, columns three to five ($O_{A}$ to $O_{C}$) indicate the number of observables of corresponding types, and columns six ($D_{l}$) and seven ($D_{q}$) yield, respectively, the number of `linear' and `quadratic'  doilies of a given type. } 
\label{tw3inw5}
\vspace*{0.4cm}
\scalebox{0.8}{
\begin{tabular}{|r|c|ccc|rr|}
\hline \hline
$T$        & $C^{-}$ & $O_{A}$ & $O_{B}$ & $O_{C}$  & $D_{l}$  & $D_{q}$    \\
\hline 
  1        & 7       & 0       & 7       & 8        &  $-$     &  81      \\
	2        & 7       & 0       & 9       & 6        &  27      & $-$     \\
\hline
  3        & 6       & 1       & 5       & 9        &  $-$     & 108      \\
\hline
  4        & 5       & 2       & 5       & 8        &  162     & $-$      \\
	5        & 5       & 2       & 7       & 6        &  $-$     & 162     \\
\hline
  6        & 4       & 3       & 5       & 7        &  $-$     & 324      \\
\hline
  7        & 3       & 0       & 9       & 6        &  9       & $-$      \\
	8        & 3       & 0       & 15      & 0        &  $-$     &  36     \\
  9        & 3       & 2       & 7       & 6        &  $-$     & 216      \\
 10        & 3       & 2       & 9       & 4        &  81      & $-$     \\	
 11        & 3       & 4       & 5       & 6        &  54      & $-$      \\
 12        & 3       & 4       & 7       & 4        &  $-$     &  81     \\
 13        & 3       & 6       & 9       & 0        &   3      & $-$      \\
 \hline \hline
\end{tabular}
}
\end{center}
\end{table}
The 27 observables of type $B$ lie on an elliptic quadric of $W(5,2)$, which 
can be defined as follows:
\begin{equation}
\mathcal{Q}^{-}_{(YYY)}(5,2):= x_1^2 + x_1 x_4 + x_4^2 + x_2^2 + x_2 x_5 + x_5^2 + x_3^2 + x_3 x_6 + x_6^2 = 0.
\label{3qdisteq}  
\end{equation}
Here, we took a coordinate basis of $W(5,2)$ in which the symplectic form $\sigma(x,y)$ is given by eq.\,(\ref{symplf}),
\begin{eqnarray*}
\sigma(x,y) =
(x_1 y_4 - x_4 y_1) + (x_2 y_5 - x_5 y_2) + (x_3 y_6 - x_6 y_3),
\end{eqnarray*}
so that the correspondence between the 63 three-qubit observables (see eq.\,(\ref{nqobs}))
\begin{eqnarray*}
{\cal S}_3 = \{G_1 \otimes G_2 \otimes G_3:~ G_j \in \{I, X, Y, Z \},~ j \in \{1, 2, 3\}\} \backslash {\cal I}_3
\end{eqnarray*}
and the 63 points of $W(5,2)$ is of the form (see eq.\,(\ref{nobspts}))
\begin{eqnarray*}
G_j \leftrightarrow (x_j, x_{j+3}),~j \in \{1, 2, 3\},
\end{eqnarray*}
taking also into account eqs.\,(13).
This special quadric $\mathcal{Q}^{-}_{(YYY)}(5,2)$, like any non-degenerate quadric, is a {\it geometric hyperplane} of $W(5,2)$. As a doily is also a {\it subgeometry} of $W(5,2)$, it either lies fully in $\mathcal{Q}^{-}_{(YYY)}(5,2)$ (Type 8), or shares with $\mathcal{Q}^{-}_{(YYY)}(5,2)$ a set of points that form a geometric hyperplane; an ovoid (Types 3, 4, 6 and 11), a perp-set (Types 1, 5, 9 and 12) and a grid (Types 2, 7, 10 and 13). One also observes that no quadratic doily shares a grid
with $\mathcal{Q}^{-}_{(YYY)}(5,2)$.

In addition to the distinguished elliptic quadric, there are also three distinguished hyperbolic quadrics in $W(5,2)$, namely:
the quadric whose 35 observables feature either two $X'$s or no $X$, 
\begin{equation}
\mathcal{Q}^{+}_{(ZZZ)}(5,2):= x_4^2 + x_5^2 + x_6^2 + x_1 x_4 + x_2 x_5 + x_3 x_6 = 0,
\label{3disthqz}   
\end{equation}
the one whose 35 observables feature either two $Y'$s or no $Y$ (see eq.\,(\ref{ndisthq})), 
\begin{equation}
\mathcal{Q}^{+}_{(III)}(5,2):= x_1 x_4 + x_2 x_5 + x_3 x_6 = 0,
\label{3disthqi}  
\end{equation}
and the one whose 35 observables feature either two $Z'$s or no $Z$,
\begin{equation}
\mathcal{Q}^{+}_{(XXX)}(5,2):= x_1^2 + x_2^2 + x_3^2 + x_1 x_4 + x_2 x_5 + x_3 x_6 = 0.  
\label{3disthqx}
\end{equation}
Accordingly, there are three distinguished doilies of Type 8, namely the ones the quadric $\mathcal{Q}^{-}_{(YYY)}(5,2)$
shares with these three hyperbolic quadrics.

Take the two-qubit doily.
Add formally to each observable, at the same position, the same mark from the set $\{X,Y,Z\}$. Pick up a geometric hyperplane in this
three-qubit labeled doily, and replace by $I$ the added mark in each observable that belongs to this geometric hyperplane.
One obviously gets a three-qubit doily. Now, there are 31 geometric hyperplanes in the doily, three possibilities $(X,Y,Z)$ to pick up a mark, and
three possibilities (left, middle, right) where to insert the mark; so there will be $31 \times 3 \times 3 = 279$ doilies created this way.
In particular,
out of the $15 \times 9 = 135$ doilies `induced' by perp-sets, 81 are of Type 10 and 54 of Type 11;
out of the $10 \times 9 = 90$ doilies `generated' by grids, 81 are of Type 12 and 9 of Type 8; finally, the
$6 \times 9 = 54$ doilies stemming from ovoids are all of the same type $9B$.
So, if we look at Table \ref{tw3inw5}, all doilies of Types 1 to 7, 27 doilies of type 8 and all doilies of type $9A$ can be regarded 
as `genuine' three-qubit guys,
9 doilies of type 8  that originate from grids (henceforth referred to as Type $8'$) and all doilies of types $9B$ to 13 can be viewed as `built from the two-qubit guy;
with Type 13 doilies being even more two-qubit-like.

\begin{figure}[t]
\centerline{\includegraphics[width=11.5truecm,clip=]{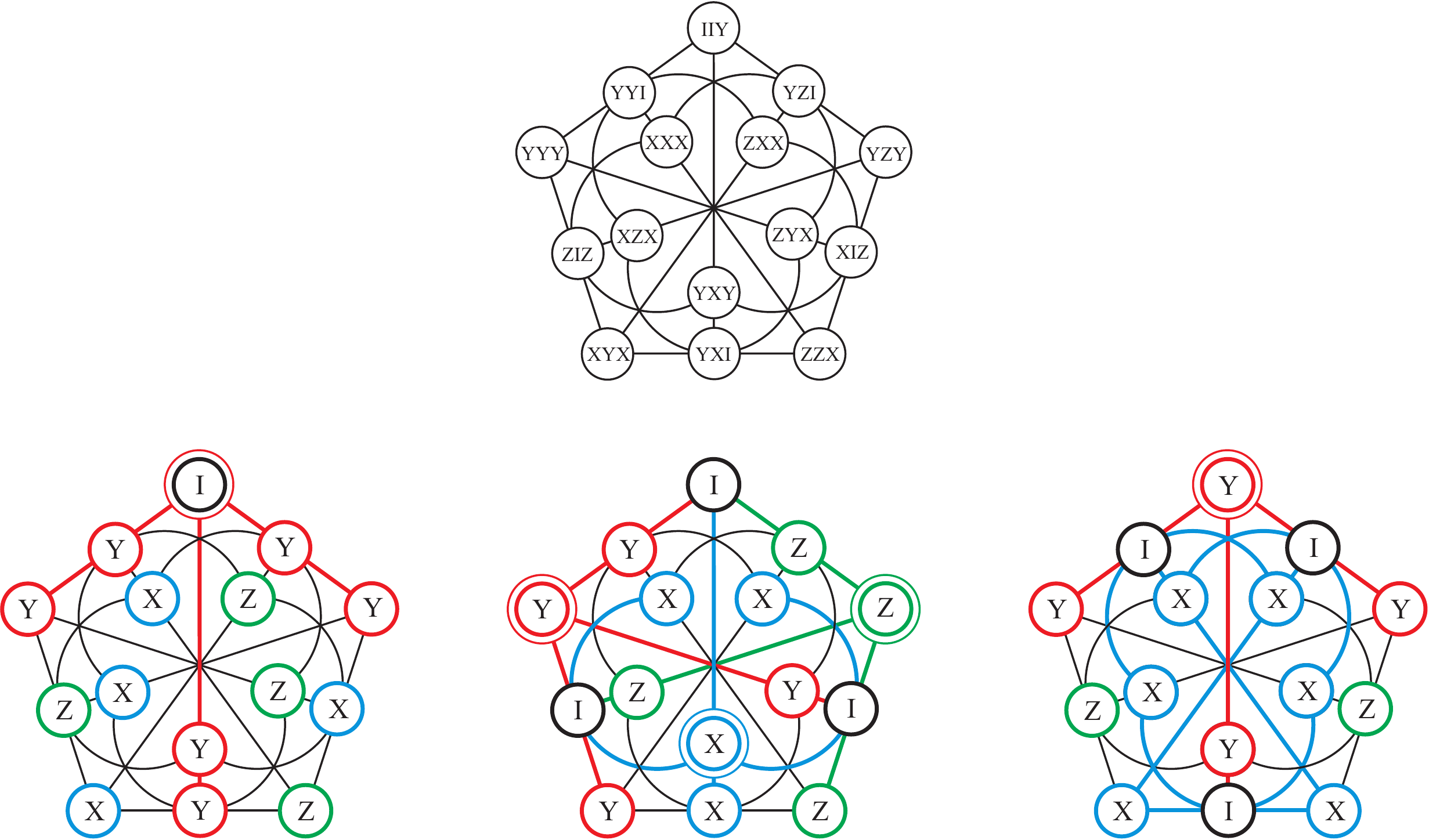}}
\vspace*{.2cm}
\caption{A formal decomposition of a three-qubit doily (top) into three `single-qubit residuals' (bottom). In each doily of the bottom row, the three geometric hyperplanes forming a Veldkamp line are distinguished by different color, with their common points being drawn black; also, the nuclei of perp-sets are represented by double circles. }
\label{type3split}
\end{figure}

This stratification of three-qubit doilies can also be spotted in a different way. Take a representative doily of a particular type, for example that of Type 3 depicted in Figure \ref{type3split}, top. From its three-qubit labels, keep first only the left mark (bottom left figure), then the middle mark (bottom middle figure) and, finally, the right mark (bottom right figure). In each of these three `residual' doilies it is easy to see that if you take the points featuring a given non-trivial mark (i.e., $X$, $Y$ or $Z$) together with the points featuring $I$, these always form a geometric hyperplane, and the whole set form a Veldkamp line of the doily where the points featuring $I$ represent its core! Employing Table \ref{t2qubvls} we readily see that this Veldkamp line is of type V (the core is a single point) for the left residual doily, type III (the core is a tricentric triad) for the middle doily and of type IV (the core is a unicentric triad) for the right one. To account this way for the 13 types of three-qubit  doilies, we also need 
the concept of a trivial Veldkamp line of the doily, i.\,e. a line consisting of a geometric hyperplane counted twice
and the full doily, which exactly accounts for those doilies `generated' by the two-qubit doily! This classification is summarized in Table \ref{trefw3inw5}. Here, columns two to six give the number of ordinary Veldkamp lines of a given type, columns seven to nine show the same for trivial Veldkamp lines and the last column corresponds to the degenerate case when all the points of a residual doily bear the label $I$. Note that all doilies stemming from the two-qubit doily (i.\,e., Types $8'$ to 13) feature ordinary Veldkamp lines of the same type. 

\begin{table}[pth!]
\begin{center}
\caption{A refined classification of doilies living in $W(5,2)$. We use the following abbreviations for the cores of Veldkamp lines: $2cl$ -- two concurrent lines, $le$ -- line, $ttr$ -- tricentric triad, $utr$ -- unicentric triad, $pt$ -- point, $ov$ -- ovoid, $ps$ -- perp-set, $gr$ -- grid and $fl$ stands for the full doily.  } 
\label{trefw3inw5}
\vspace*{0.4cm}
\scalebox{0.8}{
\begin{tabular}{|c|ccccc|ccc|c|}
\hline \hline
$T$        & 2cl & le   & ttr  & utr  & pt   & ov  & ps  & gr  & fl  \\
\hline
  1        & 1   & $-$ & $-$ & $-$ & 2   & $-$ & $-$ & $-$ & $-$    \\
  2        & $-$ & 3   & $-$ & $-$ & $-$ & $-$ & $-$ & $-$ & $-$    \\	
  3        & $-$ & $-$ & 1   &  1  & 1   & $-$ & $-$ & $-$ & $-$    \\	
	4        & $-$ & 1   & 2   & $-$ & $-$ & $-$ & $-$ & $-$ & $-$    \\
	5        & 1   & $-$ & $-$ & 2   & $-$ & $-$ & $-$ & $-$ & $-$    \\	
	6        & 1   & $-$ & 1   & 1   & $-$ & $-$ & $-$ & $-$ & $-$    \\	
	7        & $-$ & 3   & $-$ & $-$ & $-$ & $-$ & $-$ & $-$ & $-$    \\
	8        & 3   & $-$ & $-$ & $-$ & $-$ & $-$ & $-$ & $-$ & $-$    \\
	$9A$     & 1   & $-$ & $-$ & 2   & $-$ & $-$ & $-$ & $-$ & $-$    \\
	\hline
 $8'$      & $-$ & $-$ & 2   & $-$ & $-$ & $-$ & $-$ &  1  & $-$    \\
 $9B$      & $-$ & $-$ & 2   & $-$ & $-$ & 1   & $-$ & $-$ & $-$    \\
  10       & $-$ & $-$ & 2   & $-$ & $-$ & $-$ & 1   & $-$ & $-$    \\
  11       & $-$ & $-$ & 2   & $-$ & $-$ & $-$ & 1   & $-$ & $-$    \\
	12       & $-$ & $-$ & 2   & $-$ & $-$ & $-$ & $-$ &  1  & $-$    \\
\hline	
  13       & $-$ & $-$ & 2   & $-$ & $-$ & $-$ & $-$ & $-$ &  1    \\
 \hline \hline
\end{tabular}
}
\end{center}
\end{table}

\begin{figure}[t]
\centerline{\includegraphics[width=3.5cm,clip=]{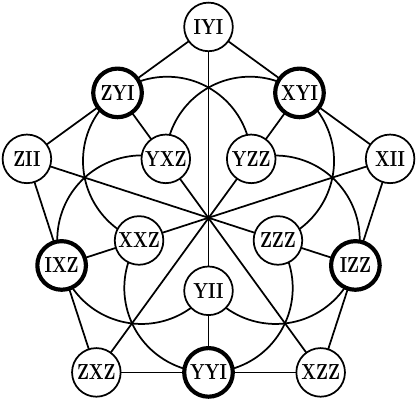}}
\vspace*{-1.3cm}
\centerline{\includegraphics[width=3.5cm,clip=]{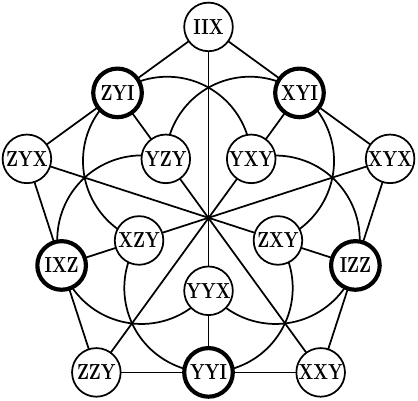}\hspace*{5.5cm}\includegraphics[width=3.5cm,clip=]{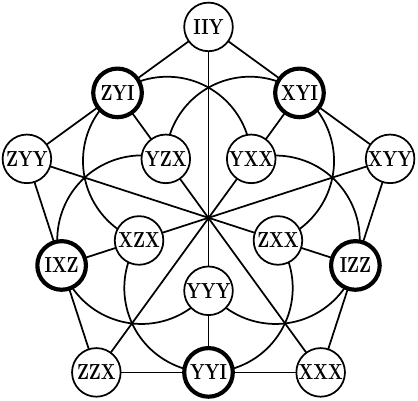}}
\vspace*{-1.3cm}
\centerline{\includegraphics[width=3.5cm,clip=]{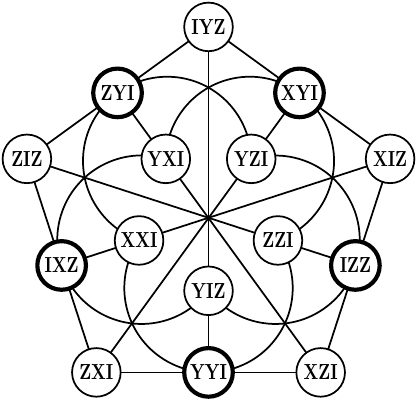}}
\caption{An illustration of the case when four different doilies share an ovoid (boldfaced). The top doily is of Type 11, the bottom one of Type 8, and both the left and right doilies are of Type 3.}
\label{4donov}
\end{figure}

Using computer, we have also found out a very interesting property that given a doily and any geometric hyperplane in it, there are three other doilies having the same geometric hyperplane. Figure \ref{4donov} serves as a visualisation of this fact when the common geometric hyperplane is an ovoid. The four doilies sharing a geometric hyperplane, however, do not stand on the same footing. This is quite easy to spot from our example depicted in Figure \ref{4donov}. A point of the doily is
collinear with three distinct points of an ovoid, the three points forming a unicentric triad.
Let us pick up such a triad, say $\{ZYI,XYI,YYI\}$ and look for its centers in each of the four doilies.
These are $IYI$ (top doily), $IIX$ (left doily), $IIY$ (right doily) and $IYZ$ (bottom one).
We see that the last three observables are mutually anticommuting, whereas the first observable
commutes with each of them. This property is found to hold for each of ${5 \choose 3} = 10$
triads contained in an ovoid. Hence, the top doily of Figure \ref{4donov} has indeed a different footing
than the remaining three. A similar $3+1$ split up is also observed in any quadruple of doilies having a grid in common 
because a point of the doily is also collinear with three points of a grid that form a unicentric triad. However, when the shared hyperplane is a perp-set, one gets a different, namely $2+2$ split, because in this case the corresponding triple of points forms a tricentric triad.

\begin{figure}[pth!]
\centerline{\includegraphics[width=12.0truecm,clip=]{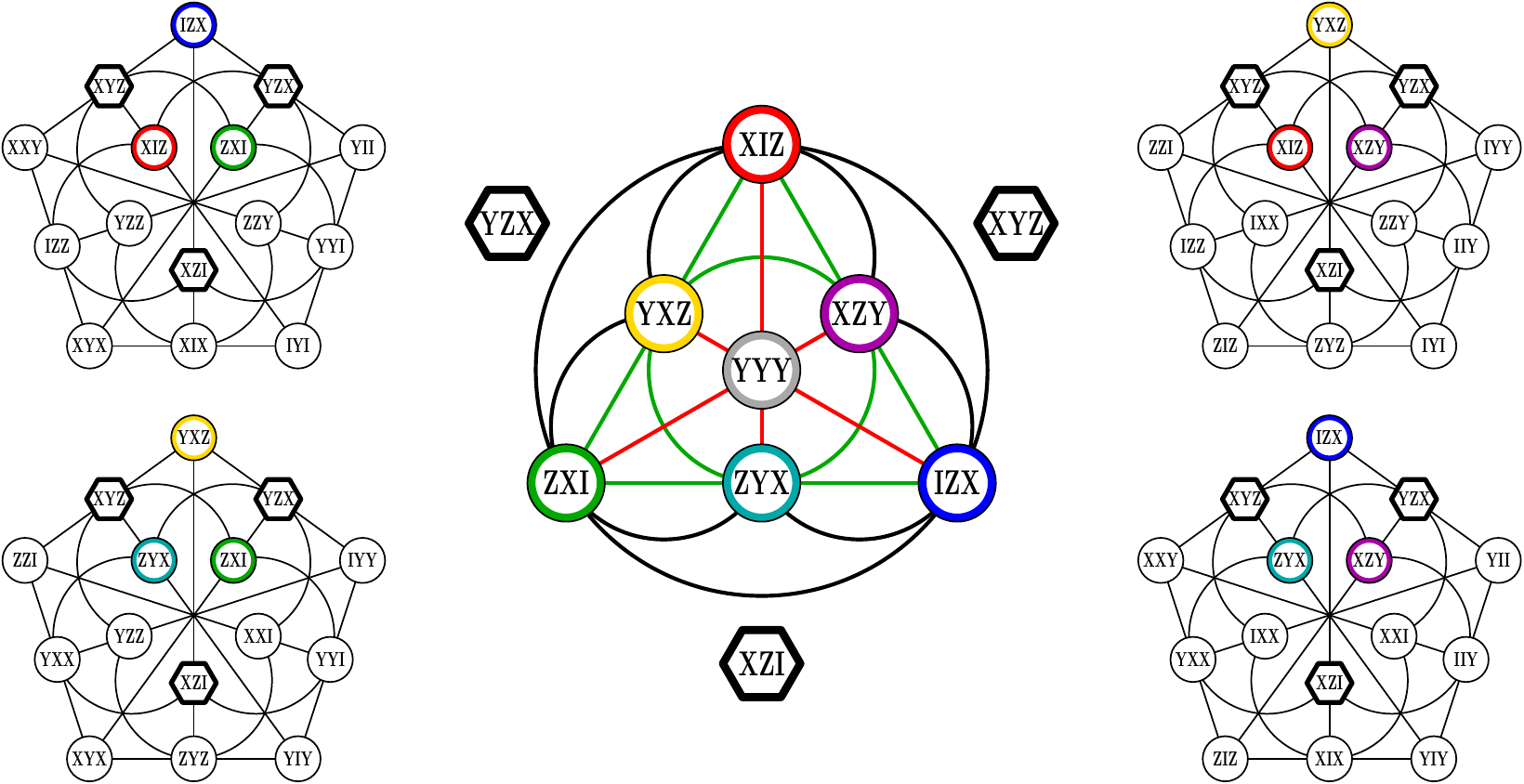}}
\vspace*{.2cm}
\caption{Four three-qubit doilies on a `planar' tricentric triad (represented by hexagons). The seven observables commuting with the three hexagonal ones are, for better illustration, colored differently. The three red lines of the Fano plane that meet at the distinguished observable (gray) are totally isotropic, whilst the remaining four (depicted green) are not. The four complementary triads (of observables) are illustrated by a full black circle and three half-circles.}
\label{quar-on-trit2}
\end{figure}

A tricentric triad of a linear resp. quadratic doily of $W(5,2)$ defines a line resp. plane in the ambient PG$(5,2)$.
The latter type of a triad is found to be shared by four quadratic doilies.
Given the three observables of such a triad, there are seven observables commuting with each of them, the corresponding seven points lying in a Fano plane (namely in the polar plane to the plane defined by the triad) in the ambient PG$(5,2)$. One of the seven observables has a distinguished footing as it commutes with each of the remaining six ones, these six observables forming three commuting pairs.
Out of the six observables one can form just four tricentric triads of which each is complementary
to the triad we started with and thus defines with the latter a unique quadratic doily. These properties are also illustrated in Figure \ref{quar-on-trit2}.

Among the 13 different types of three-qubit doilies,
there is one type, namely Type 3, which has two remarkable properties. The first property is that  
there is one point (to be called a deep point) such that all three lines passing through it are negative. Let's take a representative doily of such a type shown in Figure \ref{ill3qdoilies}, 1st row right. The deep point
is $ZIZ$. Then one sees that there are just two points (to be called zero-points) such that neither of
them lies on a negative line; one is $IIY$ and the other is $XIZ$. These two points and the deep point form
in the doily a tricentric triad, hence a copy of `linear' $W(1,2)$! The second property is related to the fact
that through each observable of type $B$ there pass four negative lines. Three of them are such that each features
one observable of type $B$ and two observables of type $C$, whereas the remaining one consists of all observables of type $B$.
Written vertically, the four negative lines passing through our deep point $ZIZ$ are:
\begin{center}
\begin{tabular}{ | c c c | c |}
 $ZIZ$ & $ZIZ$ & $ZIZ$ & $ZIZ$ \\ 
 $XXX$ & $XYX$ & $XZX$ & $XIX$ \\  
 $YXY$ & $YYY$ & $YZY$ & $YIY$  
\end{tabular}
\end{center}
We see that the three lines that are located in the doily are of \underline{the same} type, viz. $B-C-C$. If we include also the fourth negative line, viz. the $B-B-B$ one, we get what we can call a `doily with a tail.' Taking into account the above-mentioned four-doilies-per-hyperplane property, we see that there are altogether 12 doilies, four per each observable, having the same tail and all being of Type 3.

\section{`Conwell' Heptads of Doilies in $\boldsymbol{W(5,2)}$}
\begin{figure}[pth!]
\centerline{\includegraphics[width=3.5truecm,clip=]{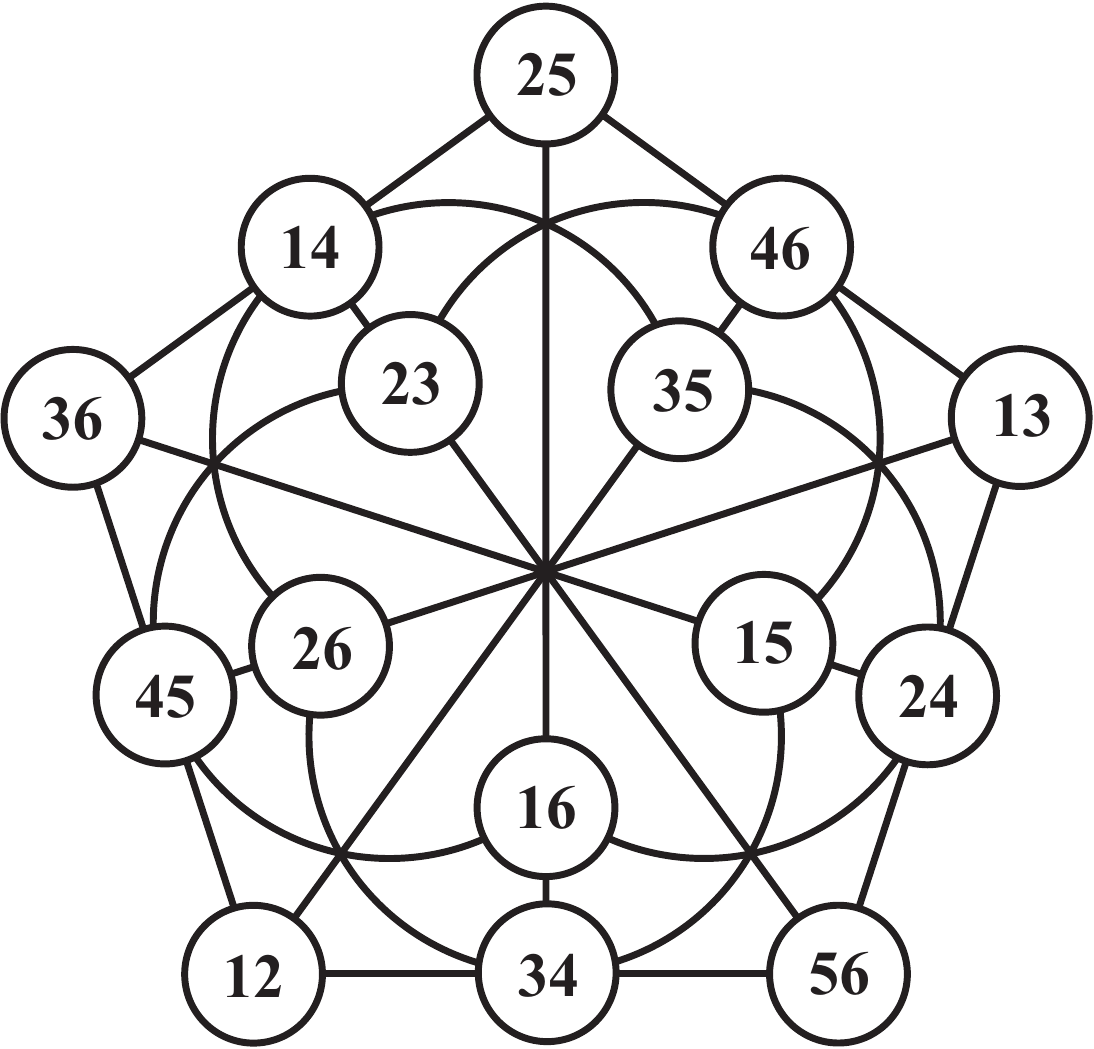}}
\vspace*{.2cm}
\caption{A duad-syntheme model of $W(3,2)$.}
\label{doily-duad-numb}
\end{figure}

\vspace*{-.4cm}
Let's recall a famous Sylvester's construction of $W(3,2)$ \cite{sylv}. Given a six-element set $M_6 \equiv \{1,2,3,4,5,6\}$,
a duad is an unordered pair $(ij) \in M_6$, $i \neq j$, and a syntheme is a set of three pairwise disjoint duads,
i.\,e. a set $\{(ij),(kl),(mn)\}$ where $i,j,k,l,m,n \in M_6$ are all distinct. The point-line incidence structure 
whose points are duads and whose lines are synthemes, with incidence being inclusion, is isomorphic to $W(3,2)$, as also illustrated in Figure \ref{doily-duad-numb}.

\begin{figure}[t]
\centerline{\includegraphics[width=11truecm,clip=]{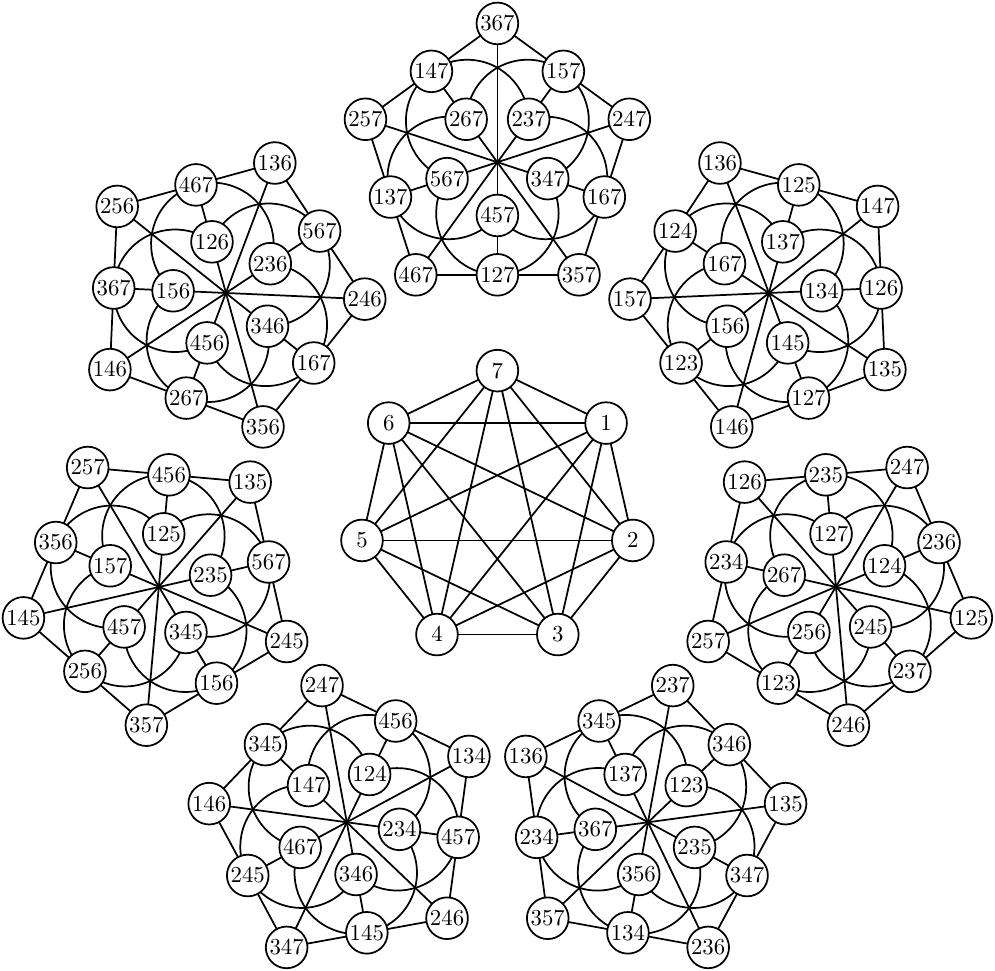}}
\vspace*{.2cm}
\caption{An abstract heptad of doilies on a seven-element set.}
\label{heptad-dusy}
\end{figure}

Next, take a seven-element set, $M_7 \equiv \{1,2,3,4,5,6,7\}$. One can form from it ${7 \choose 3} = 35$ unordered triples 
$(ijk)$, $i \neq j \neq k \neq i$. From each set of fifteen triples having the same element in common we can create a doily 
using the duad-syntheme construction on that six-element subset of $M_7$ where the common element is omitted. So, we get seven different doilies, one per each element, as depicted in Figure \ref{heptad-dusy}. Any two of them have an ovoid in common; because each ovoid is characterized by two elements, say $a$ and $b$, and it is of the form $\{(abc), (abd), (abe), (abf), (abg)\}$, where $a,b,c,d,e,f,g \in M_7$ are all different, hence it belongs to both the $a$-doily and  the $b$-doily. Also, any triple is shared by three doilies.

A remarkable fact is that this abstract heptad of doilies has a neat realization in our three-qubit $W(5,2)$. To see this, 
we have to introduce the notion of a {\it Conwell heptad} of PG$(5,2)$. Given a $\mathcal{Q}^+(5,2)$ of PG$(5, 2)$, a Conwell heptad \cite{con} (in the modern language \cite{thas}  also known
as a {\it maximal exterior set}) with respect to $\mathcal{Q}^+(5,2)$ is a set of seven off-quadric points such
that each line joining two distinct points of the heptad  is skew to the $\mathcal{Q}^+(5,2)$. There are exactly 8 heptads with respect to $\mathcal{Q}^+(5,2)$. Any two of them have exactly one point in common and any point off $\mathcal{Q}^+(5,2)$ is exactly in two heptads; also any six points of a heptad are linearly independent in PG$(5,2)$.
Next \cite{ct}, let $P$ be a point on $\mathcal{Q}^+(5,2)$. The tangent hyperplane of $\mathcal{Q}^+(5,2)$ at $P$ intersects a heptad $C$ in exactly three points $P_1,P_2$ and $P_3$ such that the points $P,P_1,P_2$ and $P_3$ are coplanar and $P_1,P_2$ and $P_3$ are not collinear; that is, the points $P_1,P_2$ and $P_3$ represent a conic in the plane and the point $P$ is its knot (the common intersection of its tangents).
Hence, there exists a bijection from the set of the 35 points of $\mathcal{Q}^+(5,2)$ onto the set of the 35 triples of points of $C$.

\begin{figure}[h]
\centerline{\includegraphics[width=11truecm,clip=]{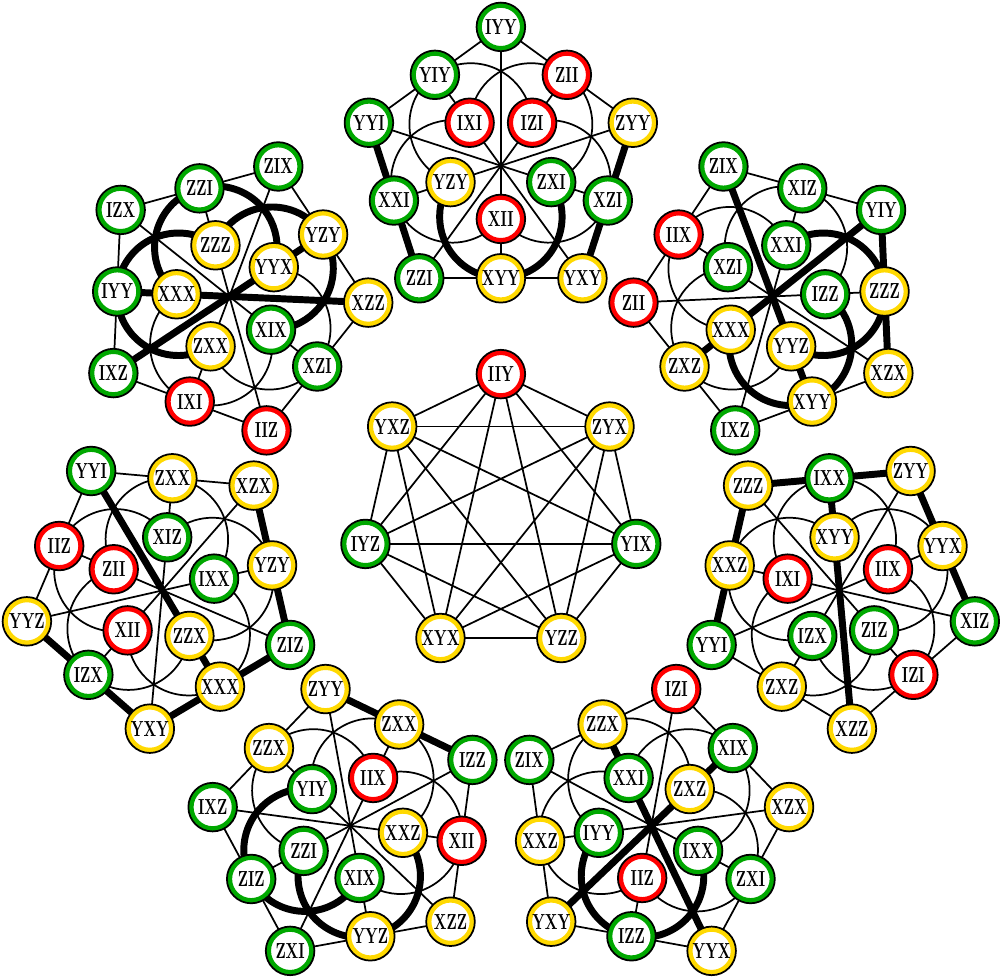}}
\vspace*{.2cm}
\caption{A `Conwell' heptad of doilies in the three-qubit $W(5,2)$. Following our convention, different types of observables are distinguished by different colors and negative lines are shown in bold.}
\label{heptad-obsv}
\end{figure}

Now, let us take a $\mathcal{Q}^+(5,2)$ that belongs to $W(5,2)$, for example $\mathcal{Q}_{(III)}^+(5,2)$ (see eq. \ref{3disthqi}) that accommodates all symmetric observables from ${\cal S}_3$.
The eight Conwell heptads with respect to this distinguished hyperbolic quadric, expressed in terms of three-qubit observables,
are:

\begin{center}
\begin{tabular}{ | c | c | c | c | c | c | c | c |}
  \fbox{1}  & \fbox{2}   & \fbox{3} & \fbox{4} & \fbox{5}   & \fbox{6}  & \fbox{7} & \fbox{8}   \\ 
 $ZYX$ & $YZI$ & $YIZ$ & $YZI$ & $YIZ$ & $YXI$ & $XYI$ & $YII$ \\ 
 $YIX$ & $YXZ$ & $YZX$ & $YXI$ & $YIX$ & $YZZ$ & $ZYZ$ & $ZYI$ \\ 
 $YZZ$ & $YXX$ & $YXX$ & $IYZ$ & $XYI$ & $YZX$ & $ZYX$ & $XYZ$ \\ 
 $XYX$ & $IYI$ & $IYX$ & $IYX$ & $IZY$ & $IYI$ & $ZIY$ & $XYX$ \\ 
 $IYZ$ & $IXY$ & $ZYZ$ & $ZIY$ & $IXY$ & $IZY$ & $XZY$ & $XIY$ \\ 
 $YXZ$ & $XZY$ & $IIY$ & $YYY$ & $YYY$ & $XXY$ & $XXY$ & $ZZY$ \\ 
 $IIY$ & $ZZY$ & $XYZ$ & $XIY$ & $$ZYI & $ZXY$ & $YII$ & $ZXY$
\end{tabular}
\end{center}
We see that each Conwell heptad entails seven pairwise anticommuting observables and so, in fact, corresponds to a set
of generators of a seven-dimensional Clifford algebra \cite{lps}.
Let us pick up one of them, say the heptad number \fbox{1}, and associate its observables with the elements of
$M_7$ as follows:
$$ 1 \leftrightarrow ZYX,  2 \leftrightarrow YIX, 3 \leftrightarrow YZZ, 4 \leftrightarrow XYX, 5 \leftrightarrow IYZ, 6 \leftrightarrow YXZ, 7 \leftrightarrow IIY.$$
From the above-described relation between tangent hyperplanes to a hyperbolic quadric and  a Conwell heptad it follows that any unordered triple $(ijk)$, $i,j,k \in M_7$, will be associated with a particular point on $\mathcal{Q}_{(III)}^+(5,2)$ and its associated observable is the (ordinary) product of the observables associated with elements/points $i$, $j$, and $k$; for example, $146 \leftrightarrow ZYX.XYX.YXZ = IXZ$. 
Hence, all seven doilies of the heptad lie fully in $\mathcal{Q}_{(III)}^+(5,2)$ and, since no two of them share a line, they partition the set of 105 lines of $\mathcal{Q}_{(III)}^+(5,2)$. Figure \ref{heptad-obsv} serves as a visualization of this particular `Conwell' heptad of doilies. As $W(5,2)$ contains 36 hyperbolic quadrics (see eq. \ref{hqinwn}), it features altogether $36 \times 8 = 288$ such heptads of doilies. 

\section{$\boldsymbol{W(7,2)}$ and its Four-Qubit $\boldsymbol{W(5,2)}$'s}
The space $W(7,2)$ possesses 255 points, 5355 lines, 11475 planes and 2295 generators, the latter being all PG(3,2)'s. Among the 255 canonical
four-qubit observables associated to the points, 12 are of type $A$, 54 of type $B$, 108 of type $C$ and 81 of type $D$. Through an observable of type $D$ there pass: four negative lines of type $D-D-D$, totaling to $\frac{81 \times 4}{3} = 108$; 12 negative lines of type $D-D-B$, totaling to
$\frac{81 \times 12}{2} = 486$; and 12 negative lines of type $D-C-C$, totaling to $81 \times 12 = 972$. Through an observable of type
$C$ there pass, apart from the above-mentioned lines of type $D-C-C$, six negative lines of type $C-C-B$, totaling to $\frac{108 \times 6}{2} = 324$. Through an observable of type $B$ there passes, apart from the already discussed two types of lines, a single negative line
of type $B-B-B$, the total number of such lines being $\frac{54 \times 1}{3} = 18$. Since no negative line can contain an observable of type $A$, the four-qubit $W(7,2)$ thus exhibits five distinct types of negative lines whose total number
is $(108+486+972+324+18=)~1908$.

\begin{table}[pth!]
\begin{center}
\caption{Classification of $W(5,2)$'s living in $W(7,2)$. Column one ($T$) shows the type, column two ($C^{-}$) the number of negative lines in a $W(5,2)$ of the given type, columns three to six ($O_{A}$ to $O_{D}$) indicate the number of observables featuring three $I$'s, two $I$'s, one $I$ or no $I$, respectively,  columns seven ($W_{l}$) and eight ($W_{q}$) yield, respectively, the number of `linear' and `quadratic' $W(5,2)$'s  of a given type, the last but one column depicts the type of intersection of a representative $W(5,2)$ with the distinguished hyperbolic quadric and the last column indicates the type of geometric hyperplane featuring the trivial mark ($I$) for composite $W(5,2)$'s.  } 
\label{tw5inw7}
\vspace*{0.3cm}
\scalebox{0.88}{
\begin{tabular}{|r|c|cccc|rr|l|c|}
\hline \hline 
$T$        & $C^{-}$ & $O_{A}$ & $O_{B}$ & $O_{C}$ & $O_{D}$ & $W_{l}$  & $W_{q}$ & Int  & GH \\
\hline
  1        & 130     & 3       & 9       & 33       & 18     &  108     &  $-$    & ell  & $---$ \\
	\hline
  2        & 126     & 0       & 24      & 0        & 39     &  $-$     &  108    & full & $---$ \\	
	3        & 126     & 1       & 13      & 27       & 22     &  $-$     &  1944   & hyp  & $---$ \\	
  4        & 126     & 2       & 10      & 30       & 21     &  $-$     &  1620   & perp & $---$ \\	
\hline
  5        & 122     & 1       & 15      & 27       & 20     &  972     &  $-$    & hyp  & $---$ \\	
  6        & 122     & 2       & 10      & 30       & 21     &  $-$     &  648    & perp & $---$ \\	
\hline
  7        & 118     & 0       & 16      & 32       & 15     &  $-$     &  324    & perp & $---$ \\	
	8        & 118     & 3       & 9       & 33       & 18     &  648     &  $-$    & ell  & $---$ \\	
  9        & 118     & 3       & 11      & 25       & 24     &  $-$     &  1296   & hyp  & $---$ \\	
\hline
 10        & 114     & 1       & 15      & 27       & 20     &  324     &  $-$    & hyp  & $---$ \\	
 11        & 114     & 1       & 17      & 27       & 18     &  $-$     &  216    & hyp  & $---$ \\
 12        & 114     & 3       & 13      & 25       & 22     &  1944    &  $-$    & hyp  & $---$ \\	
 13        & 114     & 4       & 12      & 28       & 19     &  $-$     &  1944   & perp & $---$ \\	
\hline
 14        & 110     & 3       & 15      & 25       & 20     &  $-$     &  1944   & hyp  & $---$ \\
 15        & 110     & 5       & 11      & 23       & 24     &  648     &  $-$    & hyp  & $---$ \\	
\hline
 16        & 106     & 5       & 13      & 23       & 22     &  $-$     &  1944   & hyp  & $---$ \\	
\hline
 17        & 102     & 1       & 21      & 27       & 14     &  $-$     &  648    & hyp  & $---$ \\	
 18        & 102     & 2       & 18      & 30       & 13     &  $-$     &  324    & perp & $---$ \\
 19        & 102     & 3       & 15      & 25       & 20     &  $-$     &  648    & hyp  & $---$ \\	
 20        & 102     & 4       & 12      & 28       & 19     &  $-$     &  1944   & perp & $---$ \\	
\hline
 21        & 90      & 0       & 36      & 0        & 27     &  $-$     &  12     & full & ell: $O=YYY$ \\		
 22        & 90      & 2       & 22      & 30       & 9      &  $-$     &  108    & perp & hyp: all 9 $O$'s featuring two $Y$'s\\
 23        & 90      & 3       & 9       & 33       & 18     &  36      &  $-$    & ell  & $---$ \\
 24        & 90      & 3       & 21      & 25       & 14     &  324     &  $-$    & hyp  & perp: all 27 $O$'s of type $C$ \\
 25        & 90      & 4       & 16      & 28       & 15     &  $-$     &  324    & perp & ell: all 27 $O$'s featuring one $Y$\\	
 26        & 90      & 5       & 15      & 31       & 12     &  324     &  $-$    & ell  & perp: all 27 $O$'s of type $B$ \\	
 27        & 90      & 6       & 18      & 26       & 13     &  $-$     &  324    & perp & hyp: 26 $O$'s having no $Y$ + $III$\\		
 28        & 90      & 7       & 17      & 21       & 18     &  108     &  $-$    & hyp  & perp: all 9 $O$'s of type $A$ \\	
 29        & 90      & 9       & 27      & 27       & 0      &    4     &  $-$    & ell  & full $W(5,2)$\\	
 \hline \hline
\end{tabular}
}
\end{center}
\end{table}
When it comes to $W(5,2)$'s, we find 11 types among their 5440 linear members and as many as 18 types among their 16320 quadratic cousins $-$ as summarized in Table \ref{tw5inw7}. 
It represents no difficulty to check that 54 observables of type $B$ and 81 ones of type $D$ lie on a particular hyperbolic quadric in $W(7,2)$, to be referred to as the distinguished hyperbolic quadric $\mathcal{Q}^{+}_{(YYYY)}(7,2)$, which is also a geometric hyperplane in the latter space. A $W(5,2)$ either lies fully in this quadric (Types 2 and 21) or shares with it a set of points that forms a geometric hyperplane. Hence, the sum of $O_B$ and $O_D$ in each row of Table \ref{tw5inw7} must be one of the following numbers: 27 (when the hyperplane of $W(5,2)$ is an elliptic quadric), 31 (a perp-set) and/or 35 (a hyperbolic quadric); for the reader's convenience, the type of such geometric hyperplane is explicitly listed in column 9 of Table \ref{tw5inw7}. 
One sees that no linear $W(5,2)$ shares with $\mathcal{Q}^{+}_{(YYYY)}(7,2)$ a perp-set and no quadratic $W(5,2)$
cuts this distinguished quadric in an elliptic quadric.
Comparing Table \ref{tw5inw7} with Table \ref{tw3inw5} one readily discerns that whereas $W(3,2)$'s in $W(5,2)$ are endowed with
both an even and odd number of negative lines, for $W(5,2)$'s in $W(7,2)$ this number is always even; in addition, the difference in $C^{-}$ for any two distinct types of four-qubit $W(5,2)$'s is a multiple of four. 

Let us have a closer look at $W(5,2)$'s featuring 90 (i.e., the smallest possible number of) negative lines. We can easily show that almost
all of them originate from the three-qubit $W(5,2)$. First, by adding $I$ to each three-qubit observable at the same position  we get the four trivial four-qubit $W(5,2)$'s of Type 29. Next,      
adding to each observable at the same position a mark from the set $\{X,Y,Z\}$, picking up a geometric hyperplane in this
four-qubit labeled $W(5,2)$ and replacing by $I$ the added mark of each observable in the geometric hyperplane 
one gets a four-qubit $W(5,2)$ with 90 negative lines. Now, there are 28 ($\#$ of elliptic quadrics) + 36 ($\#$ of hyperbolic quadrics) + 63 ($\#$ of perp-sets) = 127 geometric hyperplanes in the $W(5,2)$, three possibilities $(X,Y,Z)$ to pick up a mark, and
four possibilities (left, middle-left, middle-right, right) where to insert the mark. So, there will be $127 \times 3 \times 4 = 1524$ four-qubit $W(5,2)$'s created this way, which only falls short by 36 the total number of $W(5,2)$'s endowed with 90 negative lines (the four guys of Type 29 being, of course, disregarded). A concise summary is given in the last column of Table \ref{tw5inw7}, where the type of geometric hyperplane is further specified by the character/type of the associated (three-qubit) observable. One observes that Type 23 is the only irreducible type of $W(5,2)$'s having 90 negative lines.

We shall illustrate this process by a couple of examples. Let us start with the perp-set of the three-qubit $W(5,2)$ whose nucleus is an observable of type $A$, say $XII$. Out of 31 observables commuting with this observable there are 7 of type $A$
($XII$, $IXI$, $IIX$, $IYI$, $IIY$, $IZI$ and $IIZ$), 15 of type $B$ ($IXX$, $IXY$, $IXZ$, $XXI$, $XIX$, $IYX$, $IYY$, $IYZ$, 
$XYI$, $XIY$, $IZX$, $IZY$, $IZZ$, $XZI$, and $XIZ$) and 9 of type $C$ ($XXX$, $XXY$, $XXZ$, $XYX$, $XYY$, $XYZ$, $XZX$, $XZY$, and $XZZ$). Hence, out of 32 observables off the perp, there will be $9 - 7 = 2$ of type $A$, $27 - 15 = 12$ of type $B$ and
$27 - 9 = 18$ of type $C$:
\begin{center}
\begin{tabular}{ | c | c c  c |}
 $\widehat{\mathcal{Q}}_{(XII)}$ & $O_A$ & $O_B$ & $O_C$ \\ 
 on             &   7   &   15  &   9 \\  
 off            &   2   &   12  &  18  
\end{tabular}~.
\end{center}
Next, each observable of the perp-set  acquires a trivial mark $I$ and hence goes into the four-qubit observable of the same type. However, an observable lying off the perp-set gets a non-trivial label  $X$, $Y$ or $Z$ and so yields
the four-qubit observable of the subsequent type; that is, $O_A^{(3)} \to O_B^{(4)}$, $O_B^{(3)} \to O_C^{(4)}$ and $O_C^{(3)} \to O_D^{(4)}$. Hence, in our case we get:
\begin{center}
\begin{tabular}{ | c | c c  c c |}
 ($\widehat{\mathcal{Q}}_{(XII)}$)      & $O_A$ & $O_B$ & $O_C$ & $O_D$\\ 
 (on -- type intact)   &   7   &   15  &   9   &  0 \\  
 (off -- type shifted) &   0   &    2  &  12   & 18 \\
  Total                &  \encircled{7}   &   \encircled{17}  &   \encircled{21}  & \encircled{18}
\end{tabular}~.
\end{center}
Comparing with Table \ref{tw5inw7}
 we see that this is a four-qubit $W(5,2)$ of Type 28.

As the second example  we shall take the case when the geometric hyperplane of $W(5,2)$ is an elliptic quadric generated by
an antisymmetric observable of type $B$, say $YXI$. This quadric, $\mathcal{Q}_{(YXI)}^{-}(5,2)$, consists of all symmetric observables that commute with
$YXI$ and all antisymmetric observables that anticommute with $YXI$. In particular, it contains 4 observables of type $A$ ($IXI$, $IIX$, $IIZ$ {\rm and} $IYI$), 11 observables of type $B$ ($XZI$, $ZZI$, $YIY$, $IXX$, $IXZ$, $YZI$, $IYX$, $IYZ$, $XIY$, $ZIY$ {\rm and} $IZY$) and 12 observables of type $C$ ($XZX$, $ZZX$, $XZZ$, $ZZZ$, $YXY$, $XYY$, $ZYY$, $YZX$, $YZZ$, $XXY$, $ZXY$ {\rm and} $YYY$).
So, out of 36 observables off the quadric, there will be 5, 16 and 15 of type $A$, $B$ and $C$, respectively. In a succinct form,

\begin{center}
\begin{tabular}{ | c | c c  c |}
 $\mathcal{Q}_{(YXI)}^{-}(5,2)$ & $O_A$ & $O_B$ & $O_C$ \\ 
 on             &   4   &   11  &  12 \\  
 off            &   5   &   16  &  15  
\end{tabular}~.
\end{center}
From this it follows that the corresponding four-qubit $W(5,2)$ is of Type 25:
\begin{center}
\begin{tabular}{ | c | c c  c c |}
 ($\mathcal{Q}_{(YXI)}^{-}(5,2)$)   & $O_A$ & $O_B$ & $O_C$ & $O_D$\\ 
 (on -- type intact)      &   4   &   11  &   12  &  0 \\  
 (off -- type shifted)    &   0   &    5  &   16  & 15 \\
  Total                   &   \encircled{4}   &   \encircled{16}  &   \encircled{28}  & \encircled{15}
\end{tabular}~.
\end{center}

\section{Conclusion}
\vspace*{-.4cm}
We have introduced a remarkable observable-based taxonomy of subspaces of $W(2N-1,2)$, $2 \leq N \leq 4$, whose rank is just one less than that of the ambient space. Alongside the distribution of various types of observables, an important parameter of the classification was
the number of negative lines contained in a subspace. As already mentioned in the introduction, this latter parameter is essential in checking whether a given finite geometric configuration is contextual or not. For example, our preliminary analysis shows that all three-qubit and four-qubit doilies are, like their two-qubit sibling, contextual. In a separate paper we
plan to address this question in more detail, employing also the degree of contextuality, for a variety of other symplectic subspaces.
However, when approaching this way subspaces of higher rank, 
it would be natural to include as parameters the number of negative linear subspaces of every viable dimension from $1$ to $N-2$, i.\,e. consider negative lines, negative planes, \ldots, negative generators; so, already in the case of $N=4$ we can add one more parameter, the number of negative planes a four-qubit $W(5,2)$ is endowed with, to get an interesting refinement of our Table \ref{tw5inw7}. As the three-qubit $W(5,2)$ features 54 negative planes \cite{shj}, each composite four-qubit $W(5,2)$
must have the same number of negative planes; in connection with this fact it would be interesting to check whether also each irreducible four-qubit $W(5,2)$ having 90 lines (Type 23) enjoys this property. 

Another interesting extension/variation of our taxonomy would be to take into account the number of negative lines passing through a point of the subspace. Let us call this number the order of a point and for each subspace $W(2s-1,2)$ define the following string of parameters $[p_0, p_1, p_2, \ldots, p_{4^{s-1} -1} ]$, where $p_k$, $0 \leq k \leq 4^{s-1} -1$, stands for the number of points of order $k$ the subspace contains. Applying this to three-qubit doilies ($s=2$), we find the following five patterns (as readily discerned from Figure \ref{tw3inw5}):  $[0,9,6,0]$ (Types 1 and 2), $[2,9,3,1]$ (Type 3), $[5,5,5,0]$ (Types 4 and 5), $[6,6,3,0]$ (Type 6) and $[6,9,0,0]$ (Types 7 to 13).

A slightly different possibility of employing our strategy is to analyse other distinguished subgeometries of $W(2N-1,2)$ like, for example, the split Cayley hexagon of order two \cite{psm}. This generalized polygon can be embedded into $W(5,2)$, and in two different ways at that \cite{cool}, called classical and skew. We have already discerned two distinct kinds of the former, and as many as 13 different types of the latter. Yet, a full understanding of the case requires a more rigorous computer-assisted approach and will, therefore, be treated in a separate paper.

\section*{Acknowledgments}
\vspace*{-.4cm}
This work was supported by the Slovak VEGA Grant Agency, Project $\#$ 2/0004/20, the French ``Investissements d'Avenir'' programme, project ISITE-BFC (contract ANR-15-IDEX-03) and by the EIPHI Graduate School (contract ANR-17-EURE-0002). 
The computations have been performed on the supercomputer facilities of the M\' esocentre de calcul de Franche-Comt\'e; the code is freely available at  {\tt https://quantcert.github.io/Magma-contextuality/}. We are also indebted to Zsolt Szab\'o and Petr Pracna for their help with the figures and thank P\'eter Vrana for providing us with a list of doilies of $W(5,2)$.

\end{document}